\documentclass[twocolumn]{article}

\usepackage{graphicx}
\usepackage{float}
\usepackage[right=20mm,left=20mm]{geometry}
\usepackage{xcolor}
\usepackage{amsmath}
\usepackage{latexsym}
\usepackage{amssymb}
\usepackage{mathptmx}
\usepackage{array}
\usepackage{mathtools}
\usepackage{authblk}
\usepackage{subfigure}     
\usepackage[colorlinks=true,allcolors=blue]{hyperref}
\usepackage{widetext}
\usepackage{graphicx}
\usepackage{float}
\usepackage{array}

\usepackage[numbers]{natbib}

\title{Scale-dependent and background-preserving gravity from an action: cosmological tests}
\author[1,2,3]{Wiliam S. Hipólito-Ricaldi} 
\author[3,4]{Rodrigo von Marttens}
\author[2]{Felipe de Melo-Santos} 
\author[2,3,5]{Davi C. Rodrigues}

\affil[1]{Departamento de Ciências Naturais, Universidade Federal do Espírito Santo. São Mateus, ES, Brazil.}
\affil[2]{Núcleo de Astrofísica e Cosmologia, Universidade Federal do Espírito Santo. Vitória, ES, Brazil.}
\affil[3]{PPGCosmo, Universidade Federal do Espírito Santo. Vitória, ES, Brazil.}
\affil[4]{Universidade Federal da Bahia, Salvador-BA, 40170-110, Brazil.}
\affil[5]{Departamento de Física, Universidade Federal da Espírito Santo, Vitória-ES, Brazil.}

\affil[ ]{\text{wiliam.ricaldi@ufes.br}, \text{rodrigomarttens@ufba.br}, \text{davi.rodrigues@ufes.br},
\text{felipedmsantos@hotmail.com}}
\date{}                     
\begin{document}

\twocolumn[
  \begin{@twocolumnfalse}
    \maketitle
    \begin{abstract}
    We investigate the observational implications of a gravitational model wherein the gravitational constant $G$ and the cosmological constant $\Lambda$ exhibit scale-dependent behavior at the perturbative level, while preserving the General Relativity (GR) field equations at the background. This model is motivated by the potential influence of large-scale (infrared) Renormalization Group (RG) corrections to gravity and is constructed upon an effective action incorporating a scale definition via Lagrange multipliers. We explore the effects of these modifications during the recombination epoch with particular focus on their impact on the structure of acoustic oscillations. Additionally, we perform a comprehensive parameter fitting analysis using data from the Cosmic Microwave background (CMB), type Ia Supernovae (SN Ia), Baryon Acoustic Oscilations (BAO) and Redshift Space Distortions (RSD). Our results indicate that the RG corrections here considered are consistent with the main predictions of the $\Lambda$CDM model, and they slightly increase the uncertainties in the parameter estimations. Such small differences cannot be used to dismiss the current cosmological tensions. Although previous results indicated that this model is more flexible than $\Lambda$CDM regarding RSD data, potentially alleviating tensions, this advantage becomes negligible with the current extended data set. The framework maintains its theoretical consistency and foundation; however, unless further generalized, it cannot effectively address current cosmological issues.
    \end{abstract}
  \end{@twocolumnfalse}
  \vspace*{1cm}
]

\section{Introduction}

Understanding the fundamental nature of gravity stands as one of the foremost challenges in modern physics, with General Relativity (GR) being the most successful theory according to our current understanding.
Combining GR with the presence of cold dark matter and dark energy, characterized by the cosmological constant $\Lambda$, yields the standard cosmological model known as the $\Lambda$CDM. This framework 
provides a comprehensive description for the current phase of accelerated expansion of the Universe, and is well aligned with most of the current observational data available \cite{SupernovaSearchTeam:1998fmf,perlmutter1999measurements,WMAP2003,Aghanim2020,cooke2018one,Eisenstein2005}.

On the other hand, recent data have relevant discrepancies with respect to the predictions of the $\Lambda$CDM model. 
One of the most intriguing deviations comes from measurements of the Hubble parameter derived from distance measurements from local Cepheids and Type Ia supernovae (SN Ia) within the SH0ES project, indicating a value of $H_0=73.04\pm1.04$~\cite{Riess:2021jrx}. This value contrasts with the Planck estimation (TTTEEE+lowE+lensing+BAO) of $H_0=67.66\pm0.42$~\cite{Planck:2018vyg}, representing a notable discrepancy at a significance level of $\sim 5\sigma$. This inconsistency has persisted over the past few years and is currently referred to as the Hubble tension \cite{DiValentino:2021izs,Abdalla2022}. Additionally, $\Lambda$CDM faces several other challenges, such as the cosmological constant problem, the cosmic coincidence problem, and more recently, a $2$–$3\sigma$ tension in the $S_8$ parameter ($S_8 = \sigma_8 \sqrt{\Omega_m / 0.3}$) has been observed between measurements from high redshift cosmic microwave background (CMB)  and low redshift weak lensing surveys and redshift space distortions \cite{Planck:2018vyg,Skara2019,Nunes2021}. 

Overall, there is a widespread belief that a better understanding of the nature of the dark sector components might shed light on the aforementioned challenges. In this context, several alternative models have been proposed to solve, or at least alleviate, the issues faced by the $\Lambda$CDM model. Among others, a particular approach of great interest in the literature involves exploring modified theories of gravity, which depart from GR and offer alternative descriptions of gravitational phenomena. In this paper, we consider a modified gravity theory based on infrared corrections to GR derived from Renormalization Groups (RG), in which $G$ and $\Lambda$ are no longer constants (e.g., \cite{Antoniadis:1991fa, Goldman:1992qs, Bertolami:1993mh, Bertolami:1995rt, Shapiro:2004ch, Reuter:2004nx, Babic:2004ev, Shapiro:2009dh, Rodrigues:2009vf, Sola:2013fka, Sola:2019jek, Giacchini:2020zrl, Alvarez:2022wef, Bertini:2024onw}). In particular, we develop here the observational consequences of Ref.~\cite{Bertini:2019xws}, which mainly builds on the works of Refs.~\cite{Shapiro:2004ch, Reuter:2003ca, Rodrigues:2009vf, Rodrigues:2015hba}. In the latter formulation, all the dynamics come straight from an action. There are no changes to the background field equations (which are still the Friedmann-Robertson-Walker ones), but the cosmological perturbations dynamics are sensitive to the running of $G$ and $\Lambda$, leaving imprints on several cosmological processes, which are discussed below. From now on, this approach, as in Ref.~\cite{Rodrigues:2009vf}, is here denoted as RGGR, in reference to Renormalization Group improved General Relativity. Its departure from GR is quantified by the parameter $\nu$, such that $\nu = 0$ reproduces the GR results.

We employed Markov Chain Monte Carlo (MCMC) methods to test the RGGR model using Baryon Acoustic Oscillations (BAO) data from 6dF Galaxy Survey~\cite{Qin:2019axr} and BOSS DR12~\cite{Alam:2016hwk},  type Ia Supernovae from Pantheon sample, CMB data from Planck (TTTEEE+lensing) and Redshift Space Distortions data (RSD). Our statistical analysis revealed no significant deviations from zero for the parameter $\nu$, indicating that its value is of the order of $10^{-5}$. Considering that data from the CMB and LSS represent two of the most precise cosmological datasets available, and acknowledging the relative success of the $\Lambda$CDM model in describing these observations, the obtained result seems reasonable. A small $\nu$ value would indeed be required to faithfully reproduce the observed high periodicity in the CMB acoustic peaks and the positioning of the BAO scale in the two-point correlation function.

This paper is organized as follows: We begin by reviewing the theoretical framework of the RGGR model. The action, field equations, and scale setting are discussed in the next Section. In Section \ref{pert} we present the equations for perturbative dynamics, while we leave Section \ref{implications} to the cosmological implications of the model, in both: early times and late times. In Section \ref{mcmc} we discuss the MCMC analysis, and finally in Section \ref{conclusions} we present our conclusions.

\section{Theoretical framework} 

\subsection{Action}

Here we consider the same action considered in Ref.~\cite{Bertini:2019xws}. This scale-dependent proposal is an extension of the renormalization group-based approach as presented in Refs.~\cite{Reuter:2003ca, Shapiro:2004ch, Reuter:2004nx, Rodrigues:2009vf}.  In particular, the work \cite{Bertini:2019xws} considers a cosmological framework for the action-based proposal introduced in Ref.~\cite{Rodrigues:2015hba}. The central idea of the latter is to fully incorporate the nontrivial large-scale (infrared) renormalization group effects to general relativity completely at the level of a classical effective action. Contrary to previous works, the specification of the renormalization group scales is also part of the action, which is done by the use of Lagrangian multipliers. If the Lagrangian multiplers values, at the field equations level, are zero; then the field equations from the complete action approach are equivalent to the field equations in which the scale settings are \textit{ad-hoc} equations. However, in general that is not the case. 

The action considered in~\cite{Bertini:2019xws} extends the action proposal of \cite{Rodrigues:2015hba}  in the sense that it allows for an arbitrary number of RG scales. More than one RG scale within quantum field theory is not a novelty (e.g., \cite{Ford:1996yc, Steele:2014dsa}), and recently another cosmological application also used two scales \cite{Bertini:2024onw}. The cosmological application of \cite{Bertini:2019xws} requires two scales in order to find a consistent action in which both the couplings $G$ and $\Lambda$ can change only due to changes on the RG scales. That is, once the first scale was set, the second scale was derived.

The scale-dependent action for gravity here considered, with the spacetime signature $(- + +  +)$, reads \cite{Bertini:2019xws}
\begin{align}
    & S[g,\mu,\lambda,\Psi, \gamma] = S_{\rm m}[g,\Psi] + \label{RGGRaction2}\\ 
    & \frac1{16 \pi } \int \left[\frac{R - 2 \Lambda(\mu)}{G(\mu)}  \!+\! \sum_p \lambda_p \left[ \mu_p - f_p(g,\Psi, \gamma)\right]\right]\! \sqrt{-g} \,  d^4x \, ,     \nonumber 
\end{align}
where $S$ is the complete action for gravity and matter fields. All fundamental fields of action $S$ are explicitly indicated in square brackets. 
$S_{\rm m}$ is the matter action, which in general depends on the matter fields $\Psi$ and the metric. The matter fields are collectively written as $\Psi$, regardless of their nature and total number. 
The metric tensor is $g_{\alpha \beta}$. $\lambda_p$, with $p=1, 2, ...$, are Lagrange multipliers. 
Each $\mu_p$ represents a RG scale whose physical meaning is provided by the corresponding function $f_p$. The scale $\mu_p$ itself is not physically essential, since it is straightforward to eliminate it in favor of $f_p$, leading to a non-minimal matter coupling \cite{Rodrigues:2015hba}.  Although this elimination of $\mu_p$ is economical considering the number of fields in the action \eqref{RGGRaction2}, there are properties that are simpler to understand using the action \eqref{RGGRaction2}. 
The auxiliary tensor field $\gamma_{\alpha \beta}$ appears only inside $f_p$ and without derivatives. The purpose of the latter is to set the RG-effects background, as will be further detailed later.

\subsection{Field equations}
From the action \eqref{RGGRaction2}, the field equations are found by doing action variations with respect to the metric $g^{\alpha \beta}$, the Lagrange multipliers $\lambda_p$, the scales $\mu_p$, the fields $\gamma$ and the matter fields $\Psi$. Using the latter order, the field equations read
\begin{align}
	{\cal G}_{\alpha \beta} + \Lambda(\mu) g_{\alpha \beta}+ f_{\alpha \beta} &= 8 \pi G(\mu) T_{\alpha \beta} \, , \label{RGGRfieldEq1}\\[.1in]
	\mu_p &= f_p  \, , \\[.1in]
	2  \frac{\partial}{\partial {\mu_p}} \frac{\Lambda}{G} -   R \frac{\partial}{\partial {\mu_p}} G^{-1}  &=  \lambda_p \, , \label{consistence} \\[.1in]
        \sum_p \lambda_p \frac{\partial f_p}{\partial \gamma_{\alpha \beta}} &= 0\, , \label{gammaEquations} \\[.1in]
    \frac 1{16 \pi} \sum_p \int \lambda'_p \frac{\delta f'_p}{\delta \Psi}  \sqrt{-g'} \, d^4x' & = \frac{\delta S_{\rm m}}{\delta \Psi} \, , \label{FieldEquationsPsi}
\end{align}
where a prime over a field means that is it computed at the spacetime point given by $x'$, while fields without a  prime are computed at $x$. Functionals, like $S$, are never written with a prime. In the above, $\delta$ is the functional derivative, thus $\delta x'/ \delta x = \delta^{(4)}(x - x')$. For deriving eq.~\eqref{consistence}, partial derivatives could be used since it is assumed that $\Lambda$ and $G$ depend on the scales $\mu_p$ directly, without derivatives. Thus, for example, $\delta \Lambda'/\delta \mu_p = \partial \Lambda / \partial \mu_p \, \delta^{(4)}(x-x')$. Similarly, in eq.~\eqref{gammaEquations} we have also replaced functional derivatives by partial ones. The definitions of $\mathcal{G}_{\alpha \beta}, f_{\alpha \beta}$ and $T_{\alpha \beta}$ are
\begin{align}
	{\cal G}_{\alpha \beta} & \equiv  G_{\alpha \beta} +   G \square  G^{-1} g_{\alpha \beta} - G \nabla_\alpha \nabla_\beta G^{-1} \, \label{Gcaldef},\\
	f_{\alpha \beta} & \equiv   - \frac {G} {\sqrt{-g}} \sum_p \int \lambda'_p \frac{\delta f'_p}{\delta g^{\alpha \beta}} \sqrt{-g'} \,  d^4 x' \, \label{fabdef}, \\
	T_{\alpha \beta} & \equiv -  \frac 2 {\sqrt{-g}} \frac{\delta S_{\rm m}}{\delta g^{\alpha \beta}} \, ,\label{Tdef}
\end{align}
with $\square \equiv \nabla_\alpha \nabla^\alpha$.

 In many modified gravity theories, $T^\alpha_\beta$ is referred to as the energy-momentum tensor by definition (see, e.g., \cite{Fabris:2020uey} and references therein), and we follow this convention here. Nonetheless, what is conserved by the diffeomorphism invariance of $S_m$ (i.e., $\delta_\xi S_m = 0$, for an arbitrary infinitesimal vector field $\xi^\alpha$) is a different quantity, namely \cite{Bertini:2019xws}
\begin{equation}
  \int \left( \frac{1}{2}  \xi^\beta \sqrt{-g} \, \nabla^\alpha T_{\alpha \beta}   + \frac{\delta S_m}{\delta \Psi} \delta_\xi \Psi \right)d^4x  = 0\, , \label{conservation} 
\end{equation}
where $\delta S_m/\delta \Psi$ satisfies eq.~\eqref{FieldEquationsPsi}.  Although certain energy and momentum are always conserved, one finds $\nabla_\alpha T^\alpha_\beta = 0$ only if $\delta S_m/\delta \Psi = 0$, or, equivalently, if either $f_p$ is independent of matter fields $\Psi$ or if $\lambda_p = 0$ at the level of the field equations. 

\subsection{Scale setting within cosmology}

There are different possible scale settings for cosmology. Here we continue the analysis of \cite{Bertini:2019xws} in which the main scale is correlated with the cosmological perturbations, not with the cosmological background. This choice preserves the dynamical equations of the background and is a cosmological extension of the scale setting proposed in \cite{Rodrigues:2009vf} in the context of galaxies, where the scale was a function of the Newtonian potential (see also \cite{Shapiro:2004ch, Domazet:2010bk, Rodrigues:2011cq, Rodrigues:2012qm, Rodrigues:2012wk}).

We assume a flat Friedmann-Robertson-Walker (FRW) spacetime for the background, and we study the evolution of the scalar perturbations. Hence, the line element in the Newtonian gauge can be written as \cite{Ma:1995ey}
\begin{equation}
	ds^2 = -a^2(\eta) (1 + 2 \psi) d\eta^2	\!+\! a^2(\eta) (1 - 2 \phi) \delta_{ij} dx^i dx^j\, , \label{lineelement}
\end{equation}
where $\eta$ is the conformal time, $a$ is the scale factor, $\phi$ and $\psi$ are the scalar perturbations. 

The spacetime metric is here denoted as  $g_{\alpha \beta}$,
and we associate  $\gamma_{\alpha \beta}$ with the cosmological background metric. Thus, 
\begin{eqnarray}
    \gamma_{\alpha\beta} = a^2(\eta) \eta_{\alpha\beta}\, .
\end{eqnarray}

Following \cite{Rodrigues:2009vf, Domazet:2010bk, Rodrigues:2015hba, Bertini:2019xws}, in the Newtonian gauge, the first RG scale $\mu_1$ is associated with the metric perturbation ($\psi$), which provides the Newtonian potential. This choice implies that the cosmological dynamics is different in either very dense or very sparse regions; moreover,  such differences are much smoother than energy-density changes. 
More precisely, and in covariant form, for a fluid with 4-velocity $U^\alpha$, the first scale $\mu_1$ can be written as a function of a scalar $W$ introduced as follows,
\begin{align} 
		\mu_1 & = f_1(W)\, ,  \label{mu1scale} \\
		W(U, g, \gamma) & \equiv U^\alpha U^\beta (g_{\alpha \beta} - \gamma_{\alpha \beta}) \overset{*}{=}- 2 \psi \, .\label{Wdef}
\end{align}
The symbol $\overset{*}{=}$ stresses that the equality was derived in a particular coordinate system, namely the comoving one ($U^i =0$). Hence, at background level, $W=0$ since $g_{\alpha \beta} = \gamma_{\alpha \beta}$. The first scale is essentially $\psi$, as previously commented. 

To express $G$ and $\Lambda$ as a function of $W$, we introduce the constants $G_0$ and $\Lambda_0$  such that 
\begin{equation}
    G|_{W = 0} = G_0 \; \mbox{ and } \; \Lambda|_{W = 0} = \Lambda_0 \, .
\end{equation}
And since we are dealing with small perturbations, it should be possible to approximate $G(\mu_1) = G(f_1(W))$ as a power expansion on $W$, thus
\begin{equation}
	G_0 G^{-1}(W) = 1 +   \nu W  + O(W^2)\, \stackrel*= 1 - 2 \nu \psi + O(\psi^2)\, ,	\label{GPhiexp}
\end{equation}
where $\nu$ is the dimensionless parameter that sets the strenght of the running. It is such that $\nu=0$ leads to standard general relativity, that is, without running on either $G$ or $\Lambda$.

If $\mu_1$ is the single scale that depends on $\gamma_{\alpha \beta}$, then eq.~\eqref{gammaEquations} implies that $\lambda_1 = 0$. Thus, if $\mu_1$ is the only scale, one finds $\delta S_m/\delta \Psi = 0$ and, therefore, $\nabla_\alpha T^{\alpha \beta} = 0$ \eqref{conservation}. This procedure was first described in \cite{Rodrigues:2015hba}.

With the above setting, if the Ricci scalar $R$, or equivalently the trace of the energy-momentum tensor $T$, is constant at the background level, then eq.~\eqref{consistence} implies that $\Lambda$ is a function of $\mu_1$ alone. In particular, considering a background cosmological constant and gravitational constants given by $\Lambda_0$ and $G_0$ respectively, it implies that $\Lambda \propto \Lambda_0 G_0 G^{-1} + O(\nu^2)$ \cite{Rodrigues:2015hba}.

On the other hand, if $T$ is not constant at the background level, as is the standard case in cosmology, $\Lambda$ must depend on something more than $\mu_1$. As done in \cite{Bertini:2019xws}, the second scale is set to be the cosmological background value of $T$. More specifically, 

\begin{align}  
	\Lambda & \approx \Lambda_0 + \left(\Lambda_0- 4 \pi G_0 T_0     \right)\delta_G\,,  \label{L1} 
\end{align}
with $\delta_G  \equiv  G_0\,G^{-1}-1$, and where $T_0$ is the background value of $T$ (which is not a constant in general).

In Ref.~\cite{Bertini:2019xws}, starting from the perfect-fluid action, it was shown that the perfect fluid energy-momentum tensor, 
\begin{equation}
T_{\alpha \beta} = \varepsilon U_\alpha U_\beta + p (U_\alpha U_\beta + g_{\alpha \beta}) \, ,
\end{equation}
satisfies
\begin{equation}
	\nabla_\alpha T^{\alpha \beta} =  Q^\beta \, , \label{divTdiff}
\end{equation}
with: $i$) $Q^\alpha_0 = 0$ (i.e., $Q^\alpha$ is zero at background level), $ii$) if $T_0$ is constant, then $Q^\alpha = 0$, and $iii$) $Q^i = 0$ in comoving frame ($U^i = 0$).

The energy density and pressure are divided into background and perturbative contributions as
\begin{align}
    \varepsilon & = \varepsilon_0 + \delta \varepsilon \, , \nonumber \\
    p & = p_0 + \delta p \, .
\end{align}
And the density contrast is denoted as $\delta_\varepsilon \equiv \delta \varepsilon / \varepsilon_0$.

\section{Equations for perturbative dynamics} \label{pert}
At background level, $W = 0$, which implies $G = G_0$ and $\Lambda = \Lambda_0$, thus the field equations have no dependence on $\nu$ and the same form of the Friedmann equations, namely 
\begin{eqnarray} \label{background}
	&& H^2 = \frac{8 \pi  G_0}{3} \, \epsilon_0 + \frac{\Lambda_0}{3} \, , \label{friedmann1}\\
	&& H' = -4 \pi  G_0 a \, (\epsilon_0 + p_0) \, . \label{friedmann2}
\end{eqnarray} 
We consider that the universe is composed by: photons ($\gamma$), neutrinos ($n$), baryons ($b$) and cold dark matter ($c$),
each of them is described by a barotropic state equation 
\begin{equation}
    {p}_{(x)} = w_{(x)} {\epsilon}_{(x)}. 
\end{equation}
At the background level the cosmic fluids are conserved, therefore 
\begin{eqnarray}
    {\epsilon}_{0(x)}' + 3 H ({\epsilon}_{0(x)} + {p}_{0(x)}) = 0\,.
\end{eqnarray}

Indeed, from the time-time component of the field equations,
\begin{align}
    \label{poisson}
    & 3 \mathcal{H} (\phi' + \mathcal{H} \psi) + k^2 \phi +3 \nu \mathcal{H}  \psi' + \nu k^2\psi - 3 \mathcal{H}' \nu \psi  = \nonumber \\
    & \;\;\;\;\; - 4 \pi G_0 a^2 (\delta \epsilon_{(c)} + \delta \epsilon_{(b)}  + \delta \epsilon_{(\gamma)}  + \delta \epsilon_{(n)})\;,&&
\end{align}
where $\mathcal{H} \equiv a'/a$. In the above, eqs. (\ref{L1}) and (\ref{background}) were used. The traceless part of eq.~\eqref{RGGRfieldEq1} provides the anisotropic stress equation, which in Fourier space is
 \begin{eqnarray} \label{traceless}
    k^2 \left[- \phi + (1-2\nu) \psi \right] = - 12 \pi G_0 a^2 ({\epsilon}_0 + {p}_0) \sigma \,,
\end{eqnarray}
where $({\epsilon}_0 + {p}_0) \sigma = -(\hat{k}_i\hat{k}^j-\frac{1}{3}\delta^j_i)\Sigma^{i}_j$ and $\Sigma^{i}_j$ denotes the traceless component of $T^{i}_j$ \cite{Ma:1995ey}. 

The term $({\epsilon}_0 + {p}_0) \sigma$ is associated to the anisotropic stress and is caused by the quadrupolar moment distribution of the relativistic components. This term is more relevant in early times. 
Differently from the GR case, the potentials  $\phi$  and $\psi$ are never equal in late times, indeed, the gravitational slip reads \cite{Bertini:2019xws}
\begin{equation}
    \frac{\phi}{\psi}=1-2\nu \, ,
\end{equation}

From the time-space and the space-space components of the field equations in Fourier space, 
\begin{align} \label{oi}
    & k \left( \phi' + \mathcal{H} \psi + \nu \psi' - \nu \mathcal{H} \psi \right) = \\ \nonumber 
    & \;\;\;\;\; 4 \pi G_0 a^2 \left( \epsilon_c V_c + \epsilon_b V_b + \frac{4}{3} \epsilon_\gamma V_\gamma + \frac{4}{3} \epsilon_n V_n\right) \, .
\end{align}
\begin{align} \label{pressure}
    & \phi'' + 2 \mathcal{H} \phi' + \mathcal{H} \psi' + (2\mathcal{H}' + \mathcal{H}^2 )\psi + \frac{k^2}{3} (\phi - \psi) + \nonumber  \\ 
    & \;\;\;\;\;  \nu \psi'' + \nu \mathcal{H} \psi' + \frac{2}{3} \nu k^2  \psi
     - (\mathcal{H}' + 2 \mathcal{H}^2) \nu \psi =  \nonumber \\ 
     & \;\;\;\;\; 4 \pi G_0 a^2 \left( \frac{1}{3} \epsilon_\gamma \delta_\gamma +\frac{1}{3} \epsilon_n \delta_n \right)\;.
\end{align}
where $\delta p_i = c^{2}_{S,i}  \epsilon_i \delta_i $ with $c^{2}_{S} =0$ for matter and $c^{2}_{S} =\frac{1}{3}$ radiation and $V_{i}$ denotes the $i$-specie velocity. 

As in $\Lambda$CDM case,  neutrinos interact only gravitationally with the other species and baryons and photons interact via Compton process in the early universe. For relativistic species $T=0$ at the background level, thus RGGR does not change their Boltzmann equations but rather changes the gravitational potentials. Furthermore, neutrinos conserve energy separately, and photons and baryons conserve energy as a single system. Therefore eq.~(\ref{divTdiff}) turns out $\nabla_\alpha T^{\alpha \beta} = \nabla_\alpha T^{\alpha \beta}_{(c)}=  Q^\beta $. Although, in principle, baryons  could also have contribution, here we assume that only dark matter contributes to the background value of ${T}$. This is a simplifying assumption. The effects of including baryons contributions should not be large, since they are subdominant as determined, for instance, by  primordial nucleosynthesis or recombination epoch  observations. Therefore,

\begin{eqnarray}
    &&Q_{0(c)} = - \, \delta_G  \, \partial_0  \epsilon_0 = - 6 \mathcal{H} \epsilon_{0(c)} \nu \Psi\;,\label{Q0}\\
    &&Q_{i(c)} = 0\;, \label{Qi}
\end{eqnarray}
To determine $ Q_{0(c)} $ it was used $ \tilde{\epsilon}_c '+ 3 {\cal H} \tilde{\epsilon}_c = 0 $. Finally, eq. (\ref{divTdiff})  becomes 
\begin{eqnarray} \label{cdm}
    \delta'_{c} + k V_{c} - 3 \phi' &=& -6 \mathcal{H} \nu \psi \,, 
\end{eqnarray}    
\begin{eqnarray} \label{cdmV}
    \qquad V'_c + \mathcal{H} V_c -  k \sigma - k\psi = 0 \,.
\end{eqnarray}
We stress that the RGGR changes with respect to $\Lambda$CDM in the previous equations are also inside the potential $\psi$ and in its relation to $\phi$, Eqs. (\ref{poisson})–(\ref{pressure}).

\section{RGGR  cosmological implications} \label{implications}
\label{sec:rggrci}

Before a detailed statistical analysis with cosmological data, in this section we focus on a more qualitative understanding, revising and extending the analysis of \cite{Bertini:2019xws}.
Several qualitative conclusions about the RGGR effects in perturbations dynamics are found by solving the equations in the matter-dominated era, where $a\propto \eta^2$ and $\mathcal{H}=-\frac{1}{2}\mathcal{H}^2$. Furthermore, in this era, the radiation anisotropic contribution in eq.~(\ref{traceless}) is negligible, which leads to $\phi \sim (1-2\nu)\psi$, and the right side of eq.~(\ref{pressure}) is negligible. Thus, equation for the gravitational potential $\psi$ takes the form 
\begin{eqnarray} 
     (1-\nu)\psi'' + 3  (1-\nu)\mathcal{H} \psi' 
     - \frac{3}{2} \nu \mathcal{H}^2  \psi = 0 \,.
\end{eqnarray}
The solution reads \cite{Bertini:2019xws}
\begin{eqnarray} \label{solucaoPsi}
\psi = C_1 \eta^{\bar{\nu}}+C_2  \eta^{-\bar{\nu}-5}\,, 
\end{eqnarray}
where $C_1$ and $C_2$ are integration constants that may depend on $k$, and
\begin{equation}
    \bar{\nu} \equiv \frac{1}{2}\left(\sqrt{1+\frac{24}{1-\nu}}-5\right) \approx \frac{6}{5} \nu + O(\nu^2)\,.
\end{equation}
We assume henceforth $\nu < 1$. For small $|\nu|$ values, the first term is the growing mode and the second is the decaying mode. Neglecting the decaying modes, potential $\psi$ takes the simplified form
\begin{eqnarray} \label{psismallnu}
\psi \approx C_1\left(1+\frac{6}{5}\nu \ln \eta \right) \,,
\end{eqnarray}
at first order of $\nu$. Thus, it is possible to see that positive values of $\nu$ produce gravitational wells with stronger gravity, while negative values result in regions with weaker gravity, when compared with the GR case. 

At late times, anisotropic contribution to the dark matter velocity  (eq. \ref{cdmV}) is negligible and baryons  are decoupled from photons. Then, baryon velocity equation becomes the same as eq.~(\ref{cdmV}). If we focus in scales well inside of the horizon where $\psi'\approx-\mathcal{H}\psi$ and $k \gg \mathcal{H}$,  solutions for velocities take the form
\begin{equation}
    V_c \propto V_b \propto \eta^{1+\bar{\nu}} \propto a^{\frac{1}{2}+\frac{\bar{\nu}}{2}} \, .
\end{equation}
That is, peculiar velocities are affected by RGGR corrections. Furthermore,  from system (\ref{cdm})-(\ref{cdmV}) and using background equations, it is possible to find one second order equation for dark matter $\delta_c$ in late times

\begin{eqnarray} \label{cdmRGGR}
\delta''_{c} + \mathcal{H}\delta'_{c} +k^2 \psi &=& 3\psi''  +3 \mathcal{H}  \psi' + \nu \, M(\eta) 
\end{eqnarray}
with
\begin{eqnarray} \label{baryonsRGGR}
M(\eta) &=& -6\psi''  -12 \mathcal{H} \psi' - 3 \mathcal{H}^2  \psi \,,
\end{eqnarray}
while for baryons we have
\begin{eqnarray}
\delta''_{b} + \mathcal{H}\delta'_{b} +k^2 \psi &=& 3\psi''  +3 \mathcal{H}  \psi' \,.
\end{eqnarray}
It must be pointed out that, in GR, $\psi$ is roughly constant during the matter-dominated era, then the terms on the right-hand side of Eqs. (\ref{cdmRGGR}) and (\ref{baryonsRGGR}) have negligible contributions. Nevertheless, this is not the case for RGGR (see Eq. (\ref{solucaoPsi}))

The extended Poisson equation (Eq. (\ref{poisson})) in the matter-dominated era and under the sub-horizon limit takes the form 
\begin{align}
    \label{poisson2}
    k^2\psi=- 4 \pi \frac{G_0}{1-\nu} a^2 (\delta \epsilon_{(c)} + \delta \epsilon_{(b)} )\;.&&
\end{align}
Then, Eq (\ref{cdmRGGR}) in  this limit becomes:
\begin{eqnarray} \label{deltac}
\delta_c''+ \mathcal{H}\delta'_c - 4 \pi \frac{G_0}{1-\nu} a^2 (\epsilon_{(c)}\delta_{(c)}  + \epsilon_{(b)} \delta_{(b)} )=0 \,.
\end{eqnarray}

An equivalent  equation for $\delta_b$ can also be found, that is,
\begin{eqnarray} \label{deltab}
\delta_b''+ \mathcal{H}\delta_b - - 4 \pi \frac{G_0}{1-\nu} a^2 (\epsilon_{(c)}\delta_{(c)}  + \epsilon_{(b)} \delta_{(b)} )=0 \,.
\end{eqnarray}  

The solutions are
\begin{eqnarray}
\delta_c \propto \delta_b \propto \eta^{2+\bar{\nu}}  \propto a^{1+\frac{\bar{\nu}}{2}} \,,
\end{eqnarray}
where only the growing modes were considered. The $k^2$ factor was absorbed by the proportionality constant.  For small $\nu$, the growth is suppressed for negative $\nu$ and enhanced by positive $\nu$, with respect to GR. In particular, for small scales and positive $\nu$, the RGGR effects reduce the Jean's scale $\lambda_J$, enhancing the colapse of structures \cite{Bertini:2019xws}.

To see the RGGR effects on the CMB anisotropies during recombination, equations  for the relevant multipoles $\Theta_l$ are developed below. We start by expanding the temperature fluctuations $\delta T/T$  in multipolar moments, 
\begin{equation}
  \delta T/T \equiv \Theta = \sum_{l} (-i)^l\Theta_l P_l(cos \theta) \, ,
\end{equation} 
where $P_l$ are Legendre polynomials. Within the tight-coupling approximation, where the acoustic oscillations arise, $\Theta_l = 0$ for all $l\geq 2$. Thus, the relevant moments are the monopole $\Theta_0$ and the dipole $\Theta_1$ \cite{Hu1995A}. The monopolar component   satisfies the damped-forced harmonic oscillator equation \cite{Hu1995A,Hu1995}
\begin{eqnarray} \label{monopolo}
\Theta_0''+ \frac{R'}{1+R}\Theta_0'+\frac{k^2}{3} \Theta_0 = F \,,
\end{eqnarray}
where $R=4\bar{\epsilon}_b/\bar{\epsilon}_\gamma$ is the baryon-photon ratio and \begin{eqnarray}
F= -\frac{k^2}{3} \psi -\phi'' - \frac{R'}{1+R}\phi'\,.
\end{eqnarray}
The expression above does not depend explicitly on $\nu$, it differs from the GR case only on the $\phi$ and $\psi$ solutions, which depend on $\nu$. 

For the adiabatic case,  (\ref{monopolo})  has as solution \cite{Hu1995A,Hu1995},
\begin{align}
& \Theta_0 (\eta_*)   = A_1 \cos [k\,r_s(\eta_*)] + \nonumber \\  & \frac{A_2}{k}\int^{\eta_*}_{\eta_0} d\eta' [1+R(\eta')]^{3/4} \sin[r_s(\eta_*)-r_s(\eta')]F(\eta')\,,
\end{align}
where 
\begin{equation}
r_s (\eta) =\frac{1}{\sqrt{3}}\int^\eta_0 \frac{d\eta}{\sqrt{1+R}} ,
\end{equation}
and $\eta_{*}$ stands the last scattering's time. 
While in  GR, during the matter-dominated era $\psi = \phi \approx C_1=constant$ and $F\approx -\frac{k^2}{3}C_1$ is also constant, in RGGR case neither $\phi = (1-2\nu)\psi$ or $F$ are not longer constants. 

At this point, we are  interested in a qualitative understanding on the GR and RGGR differences. For simplicity, we set  $R=0$ as a first approximation. Then, at first order in $\nu$, eq. (\ref{monopolo}) for RGGR takes the form

\begin{eqnarray}
(\Theta_0+C_1)''+ \frac{k^2}{3} (\Theta_0+C_1) =-\frac{2\,k^2}{5}C_1\,\nu  \ln \eta  \,,
\end{eqnarray}
where it was used solution (\ref{psismallnu}) and the decaying modes in $F$ were neglected. The adiabatic solution to this equation reads
\begin{eqnarray}\label{monopole}
    \Theta_0 = A_1  \cos \left(k\,r_s\right) + \nu \frac{6\,A_2}{5}  {\cal{K}}\left(k\,r_s\right) \,,
\end{eqnarray}
where $A_1$ and $A_2$ are constants and 
\begin{eqnarray}
 {\cal{K}}(x) \equiv \cos(x)\, \mbox{ci}(x) + \sin(x)\, \mbox{si}(x) - \ln\left(\frac{\sqrt{3}x}{k}\right) \, , \\
 \mbox{si}(x) \equiv \int^{x}_{0} \frac{\sin t}{t} dt \, ,  \qquad \mbox{ci}(x) \equiv -\int^{\infty}_{x} \frac{\cos t}{t} dt \, .   \nonumber
\end{eqnarray}
 The case where $R\neq 0$ enhances the compression and suppresses the rarefaction stage of the acoustic oscillation (baryon drag effect) \cite{Hu1997}, thus instroducing a modulation in the amplitudes in eq. (\ref{monopole}). The dipole can be found by using $\Theta'_0+k\Theta_1 = -\phi '$ \cite{Hu1995}.

\begin{figure}[h!]
\center
\subfigure{\includegraphics[width=0.5\textwidth]{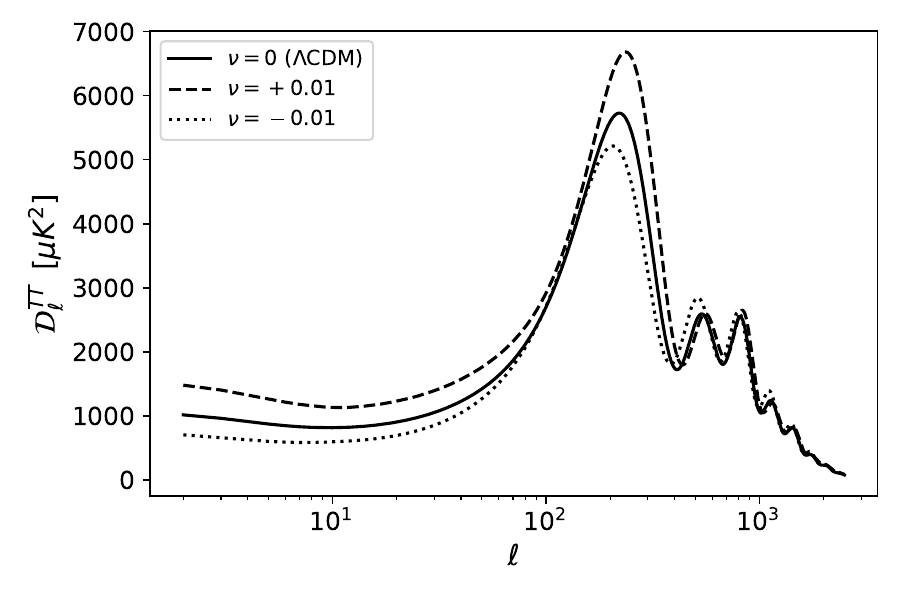}}
\caption{CMB angular power spectrum for GR ($\nu=0$), and RGGR case with $\nu=0.01$, and $\nu=-0.01$.}
 \label{fig:cmb}
\end{figure}

Qualitatively, the acoustic behavior in the CMB angular power spectrum is dominated by the features of the monopole and dipole. In the GR case ($\nu=0$), $\Theta_0$ and $\Theta_1$ are periodic, and the positions of the acoustic peaks are given by the condition $kr_s = n\pi$, with $n = 1, 2, 3, \dots$. However, this is not the case for RGGR solutions. The effect of the $\nu$ parameter in solution (\ref{monopole}) is twofold: it changes the amplitude and periodicity of $\Theta_0$ and $\Theta_1$; thus modifying the positions and amplitudes of the acoustic peaks at the same time. This effect cannot be mimicked simply by changing $\Omega_c$ or $\Omega_k$.

RGGR also leads to effects at the largest scales where the ordinary Sachs-Wolfe effect is dominant. At such scales, the angular power spectrum takes the form \cite{Dodelson2003}
\begin{eqnarray} \label{dell}
D_\ell  \propto \ell (\ell+1))\int^\infty_0 \frac{dk}{k} j^2_l(k\eta_0)\psi(\eta_0) \,, 
\end{eqnarray}
where a scale-invariant primordial spectrum was used, $j_l$ represents spherical Bessel functions, $k$ denotes the Fourier modes and $\eta_0$ indicates the current conformal time. Integrating eq. (\ref{dell})  in the GR case yields an almost  constant $D_\ell$ (a low-scales plateau). 
Clearly, from eq. (\ref{psismallnu}), we observe that RGGR causes an upward or downward shift of this plateau relative to GR, due to the term proportional to $\nu$. More quantitatively, RGGR effects on the CMB can be seen by numerically solving the complete RGGR Boltzmann system, including the baryon effects ($R \neq 0$), with the initial conditions found in Appendix \ref{initialconditions}. In  Fig. \ref{fig:cmb}, it is shown the CMB power spectra for three different $\nu$ values: $-0.01$, $0$, and $0.01$ found using CLASS. It is straightforward to see that for $|\nu| \gtrsim {\cal{O}}(10^{-2})$ the differences in the CMB temperature power spectrum are already considerable. In this context, the effects of RG are expected to be small.

\section{Statistical analysis} \label{mcmc}
\label{sec:analise}

Here we perform an analysis using the most recent public available data, the full CMB data from Planck 2018 (TTTEEE+lensing) \cite{Aghanim:2018eyx}, Type Ia Supernovae (SNe Ia) from Pantheon~\cite{Pan-STARRS1:2017jku} and Baryonic Acoustic Oscillation (BAO) from SDSS (BOSS DR12)~\cite{BOSS:2016wmc} and Redshift Space Distortions data (RSD). We aim to constraint the feasible range of $\nu$, and to disclose its degenerancies with the other cosmological parameters. Furthermore, we also assess whether the RGGR approach can have an impact on the $H_0$ and $\sigma_8$ tensions~\cite{Perivolaropoulos:2021jda}. 

The parameter space  considered here is given by the usual six $\Lambda$CDM parameters \{$\omega_{cdm}$, $\omega_{b}$, $n_s$, $H_0$, $\sigma_8$, $\tau_{\rm reio}$\} together with the RGGR parameter $\nu$. They are, respectively, the physical CDM density, $\omega_{b}\equiv\Omega_{b}\ h^2$,  the physical baryon density, the scalar spectral index $n_s$, the Hubble constant $H_0$, the amplitude of mass fluctuations $\sigma_8$,  and the reionization optical depth $\tau_{\rm reio}$. In order to compute the theoretical predictions for the cosmological observables, we used a modified version of the Boltzmann solver code CLASS~\cite{Blas:2011rf}, in which the perturbative equations found in Section ~\ref{pert} and initial conditions found in Appendix ~\ref{initialconditions} were implemented. For the statistical analysis we have used the numerical code MontePython~\cite{Brinckmann:2018cvx}. 

\subsection{Observational data}
\label{ssec:data}

Here we detail further the considered observational data in our statistical analysis.\\

\noindent
\textbf{CMB (Planck 2018)}: We utilize the latest release of Cosmic Microwave Background (CMB) data from Planck 2018 \cite{Aghanim:2018eyx}, encompassing temperature, polarization, temperature-polarization cross-correlation, and lensing reconstruction maps. Specifically, we employ the Commander likelihood code for the temperature-temperature (TT) spectrum over the range ($2 \leq \ell < 30$), and the SimAll likelihood code for the E-mode polarization (EE) spectrum with ($2 \leq \ell < 30$). For higher multipoles, we use the Plik likelihood code for the TT spectrum with ($30 \leq \ell < 2500$), and for both the temperature-polarization TE and EE spectra with ($30 \leq \ell \leq 2000$). Additionally, for the lensing spectrum, we utilize the likelihood derived from the power spectrum reconstruction spanning ($8 \leq L \leq 400$). \\

\noindent
\textbf{Type Ia Supernovae (SN Ia):} For our analysis of type Ia supernovae, we used the Pantheon sample, an extensive compilation comprising 1048 SNe Ia light curves and spectra gathered from various surveys~\cite{Pan-STARRS1:2017jku}. This information is then used for determining the distance modulus of individual SNe Ia, defined as the difference between their apparent and absolute magnitudes,
\begin{equation} \label{eq:mu}
    \mu\equiv m-M=5\log_{10}\left[\frac{d_{L}\left(z\right)}{1\,{\rm Mpc}}\right]+25\,,
\end{equation}
where $d_{L}\left(z\right)$ is the luminosity distance, expressed as,
\begin{equation} \label{eq:Dl}
    d_{L}\left(z\right)=c\left(1+z\right)\int_0^z\frac{d\tilde{z}}{H\left(\tilde{z}\right)}\,.
\end{equation}
In our analysis, the data consists of the apparent magnitudes of the SNe Ia $m$, while the absolute magnitude $M$ is treated as a nuisance parameter.

\noindent
\textbf{Baryon Acoustic Oscillation (BAO):} We utilized BAO data from the 6dF Galaxy Survey \cite{Qin:2019axr} and BOSS DR12 \cite{Alam:2016hwk}. Critical physical quantities for the BAO analysis include the sound horizon at the drag epoch of the primordial photon-baryon plasma, the angular diameter distance, and the dilation scale \cite{Eisenstein:2005su}. These quantities are defined as follows,
\begin{align}
    & r_s = \int^{\infty}_{z_{drag}} \frac{c_s (z)}{H(z)} \mbox{d}z \,, \nonumber\\ 
    & d_A = \frac{1}{1+z} \int^{z}_0 \frac{\mbox{d}z'}{H(z')}\, ,  \\  
    & D_{\mbox{v}} = \left[ (1+z)^2 d_A \frac{z}{H(z)} \right]^{1/3} \, ,\nonumber
\end{align}
where $c_s$ represents the speed of sound in the primordial photon-baryon plasma. In our model, the background dynamics remain unaffected, allowing us to derive the sound speed using the standard $\Lambda$CDM expression, 
\begin{equation}
    c_s=\frac{1}{\sqrt{3\left(1+\frac{3\Omega_{\rm b}}{4\Omega_{\rm r}}\right)}}\,.
\end{equation}

\noindent
\textbf{Redshift Space Distortions (RSD):} Similar to BAO, the RSD effect serves as a crucial tool for probing the large-scale structure of the universe, providing measurements of $f \sigma_8 (a)$. This is achieved by analyzing the ratio of the redshift space power monopole to the quadrupole, which depends on $\beta = f / b$, where $f$ is the growth rate and $b$ is the bias, as described by linear theory \cite{Song:2008qt, Percival:2008sh, Nesseris:2006er}. It is important to note that $f \sigma_8 (a)$ is independent of the bias. According to Ref.~\cite{Song:2008qt}, RSD is an effective discriminator among dark energy models and, since it is sensitive to the linear perturbations, it is an important probe in our context. Typically, a fiducial model is employed to convert angles and redshift into distances, however, the correction for the fiducial model is implemented following the approach detailed in ref.~\cite{Skara:2019usd}.
\begin{eqnarray} \label{eq:corr}
    q(z_i) = \frac{H(z_i) d_A(z_i)}{H^{fid}(z_i) d^{fid}_A(z_i)}\, ,
\end{eqnarray}
where $i = 1, \ldots, N$ (with $N$ representing the total number of data points), and the superscript ``$fid$'' signifies that the fiducial cosmology is applied. The specific parameters employed in the fiducial cosmology for each data point are detailed in Table VI of Ref.~\cite{Skara:2019usd}. The correction outlined in eq.~\eqref{eq:corr} is crucial for defining the likelihood in our statistical analysis. To construct the $\chi^2_{f \sigma_8}$ for the RSD data, we begin by defining the vector that represents the difference between observed data and theoretical predictions,
\begin{eqnarray}
    V_i \equiv f\sigma_{8,i} - \left[f\sigma_8(z_i)\right]_{\mbox{\tiny theoretical}} \, ,
\end{eqnarray}
where $f\sigma_{8,i}$ represents the value of the $i$-th data point, and $\left[f\sigma_8(z_i)\right]_{\mbox{\tiny theoretical}}$ denotes the theoretical prediction. It is important to note that the theoretical prediction consists of the RRGR computation of $f \sigma_8(a)$, divided by the correction term,
\begin{equation}
    \left[f\sigma_8(z_i)\right]_{\mbox{\tiny theoretical}} = \frac{\left[f\sigma_8(z_i)\right]_{\mbox{\tiny RGGR}}}{q(z)}\,.
\end{equation}

The likelihood function by means of a chi-squared is then built as,
\begin{eqnarray}
    \chi^2_{f \sigma_8} = \sum_{i,j} V_i C^{-1}_{ij} V_j \, .
\end{eqnarray}
For the covariance matrix $C$, we assume, as usual, that most data are not correlated, with the exception of WiggleZ data \cite{Blake:2012pj}\footnote{While the SDSS and WiggleZ survey regions partially overlap, Ref.~\cite{Skara:2019usd} demonstrates that ignoring this correlation has a negligible impact on our analysis.}. Thus, the covariance matrix is constructed using the WiggleZ covariance matrix as a block, complemented by incorporating the squared errors of the other elements along the diagonal,
\begin{displaymath}
C = \left(\begin{array}{cccc}
\sigma^2_1 & 0 & 0 & \cdot\cdot\cdot\\
0 & C_{\mbox{\tiny WiggleZ}} & 0 & \cdot\cdot\cdot\\
0 & 0 & \cdot\cdot\cdot & \sigma^2_N
\end{array}\right)\,,
\end{displaymath}
with
\begin{displaymath}
C_{\mbox{\tiny WiggleZ}} = 10^{-3}\left(\begin{array}{ccc}
6.400 & 2.570 & 0.000\\
2.570 & 3.969 & 2.540\\
0.000 & 2.540 & 5.184
\end{array}\right)\,.
\end{displaymath}

Although the RSD dataset employed in this study is extensive, it is not the most recent available. As elaborated in Sec.~\ref{ssec:results}, its constraining power on our model is limited,  our results are predominantly shaped by the analysis of CMB data.

\subsection{Numerical methods and results} \label{ssec:results}

The statistical analysis is then performed using the data presented in Sec.~\ref{ssec:data}, implemented on the MontePython code~\cite{Audren:2012wb,Brinckmann:2018cvx}, which consists in a numerical code to perform Bayesian analysis in the cosmological scenario. In our analysis, we assessed the convergence of the chains by employing the Gelman-Rubin convergence parameter $\hat{R}$ \cite{Gelman:1992zz} and requiring that $\hat{R} - 1 < 0.01$.

In our analysis, we consider a total of seven cosmological parameters\footnote{This includes nuisance parameters such as the absolute magnitude of SN Ia and additional nuisance parameters in the high-$\ell$ CMB data}: the six standard cosmological parameters from the $\Lambda$CDM model and the RGGR parameter $\nu$. It is important to note that the new dynamics introduced by the RGGR model are primarily manifested in data with only perturbative nature, specifically the CMB and RSD data. Nevertheless, the background data (BAO and SN Ia) play a crucial role in constraining the usual cosmological parameters, which may exhibit correlations with the RGGR parameter $\nu$, and therefore these background datasets are significant for indirectly constraining $\nu$.

We use a modified version of the Boltzmann solver code CLASS \cite{Blas:2011rf}. The key equations implemented in the code are Eqs. (\ref{traceless}), (\ref{oi}), and (\ref{cdm}).  The initial conditions were calculated in the RGGR context and implemented in the code. A detailed derivation of the initial conditions is provided in Appendix \ref{initialconditions}.\footnote{More specifically, when using the Newtonian gauge, the modifications to the Einstein equations are implemented starting at line 6500 of the file \texttt{class\_public/source/perturbations.c}. For the CDM conservation equations, the changes begin at line 9128 of the same file.}

\begin{figure}
    \begin{center}
        \includegraphics[width=\columnwidth]{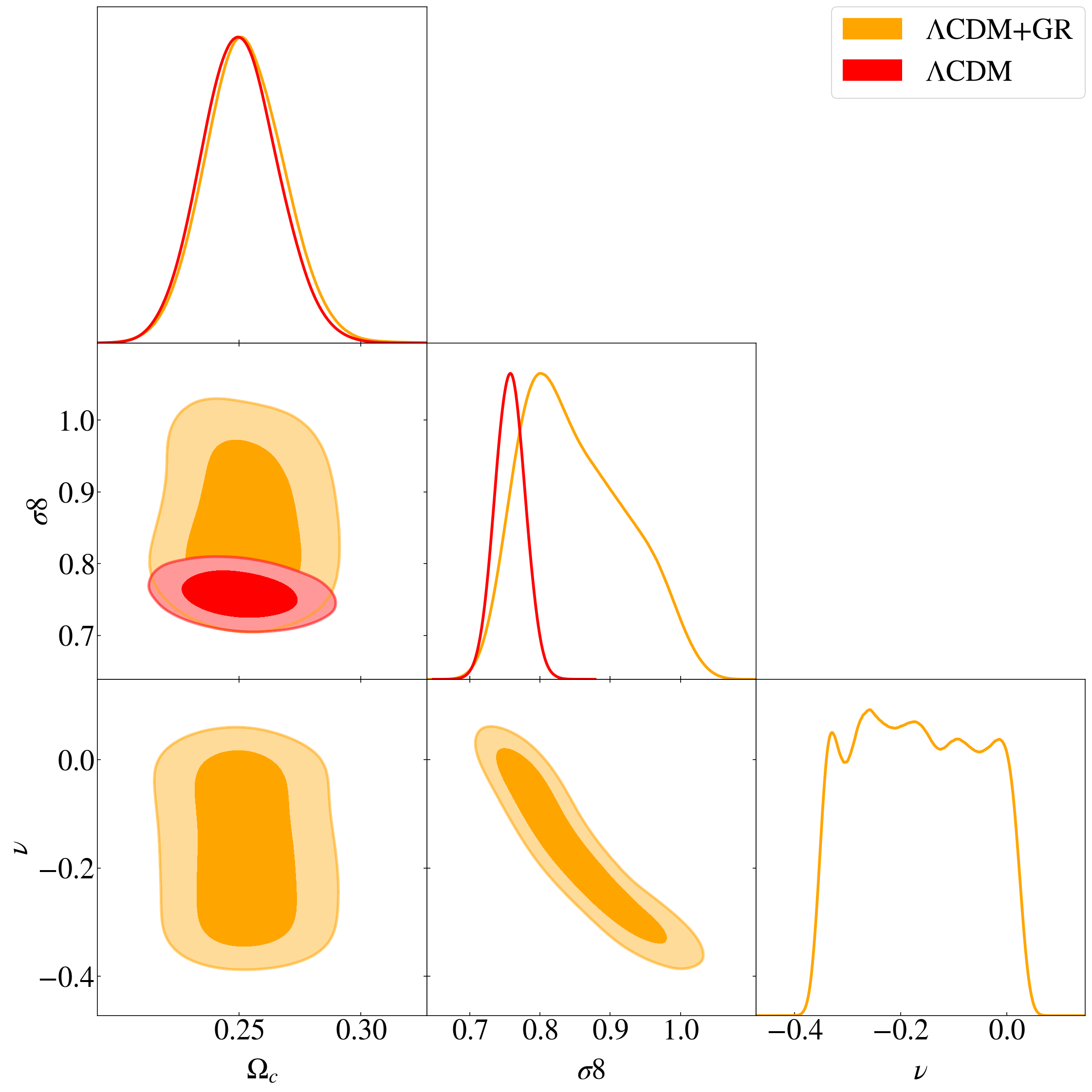}
        \caption{Analysis with data from RSD. In red we have the results for $\Lambda$CDM ($\nu=0$) and in orange we have the results for RGGR ($\nu$ is free to vary).}
        \label{plotrsd}
    \end{center}
\end{figure}

\begin{figure*}
    \begin{center}
        \includegraphics[height=15cm]{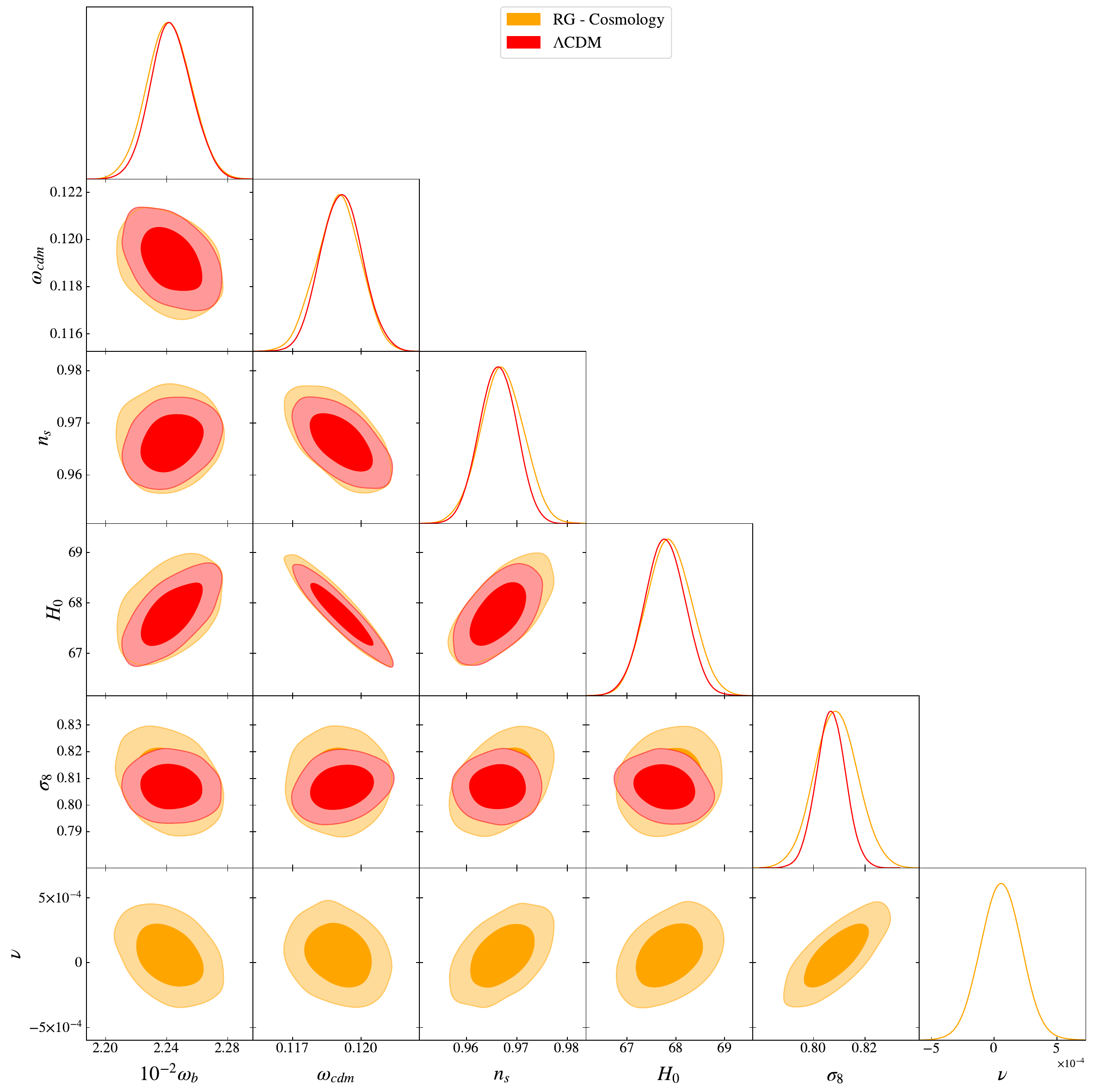}
        \caption{Statistical analysis of the joint analysis with data from CMB, BAO, SNIa and RSD. In red we have the results for $\Lambda$CDM ($\nu=0$) and in orange we have the results for RGGR ($\nu$ is free to vary).}
        \label{plotplanckbaorsd}
    \end{center}
\end{figure*}

Initially, we focus exclusively on CMB data, as presented at the top of Table~\ref{tab:results}. This analysis reveals that CMB data significantly constrain RGGR, yielding results that closely align with the $\Lambda$CDM  ($\nu \sim 10^{-5}$). Then we consider RSD data only. Results, depicted in Fig.~\ref{plotrsd},  indicate that RSD data can not strongly constrain the RGGR parameter
Subsequently,  we incorporated  data from BAO, SN Ia to CMB and RSD and perform a joint analysis CMB+BAO+SNIa+RSD. Findings from this analysis result in even stronger constraints and are summarized in Table~\ref{tab:results} and Fig.\ref{plotplanckbaorsd}. However, the results are primarily driven by CMB data. The best fit remains consistent with the $\Lambda$CDM model ($\nu = 0$) within the 1$\sigma$ confidence level. These strong constraints on the RGGR parameter arise primarily from its influence on the CMB spectral structure (see Sec.~\ref{implications}), as even small values of $\nu$ lead to substantial changes in the CMB anisotropy temperature spectrum, as demonstrated in Fig.~\ref{fig:cmb}.

The main differences between the RGGR model and the $\Lambda$CDM model are a slight shift in $H_0$ and a small increase in the error bars for the parameters $n_s$, $H_0$, and $\sigma_8$. However, in all cases, these differences are less than $1\%$ and can be considered negligible. Therefore, we conclude that RG effects are not a viable solution for addressing the $H_0$ tension or the $\sigma_8$ tension. These results are also illustrated in Fig.~\ref{plotplanckbaorsd}.
\begin{table}[H]
\caption{Result of the statistical analysis with $2\sigma$ CL for $\nu = 0$ and $\nu$ free with CMB alone and the joint CMB+BAO+SN IA+RSD analysis.}
\begin{tabular}{lll}
\hline\hline
\multicolumn{3}{c}{CMB only ananlysis}                                                            \\ \hline\hline
Parameter         & $\Lambda$CDM                  & RGGR                                          \\ \hline
$100~\omega_{b }$ & $2.237_{-0.015}^{+0.015}$     & $2.234_{-0.015}^{+0.015}$                     \\ 
$\omega_{cdm }$   & $0.1202_{-0.012}^{+0.012}$    & $0.1197_{-0.014}^{+0.014}$                    \\ 
$n_{s }$          & $0.9661_{-0.0042}^{+0.0042}$  & $0.9665_{-0.0053}^{+0.0053}$                  \\ 
$\tau_{reio }$    & $0.0548_{-0.0073}^{+0.0073}$  & $0.0552_{-0.0076}^{+0.0076}$                  \\ 
$H_0$             & $67.26_{-0.54}^{+0.54}$       & $67.58_{-0.66}^{+0.66}$                       \\ 
$\sigma_8$        & $0.8117_{-0.0060}^{+0.0060}$  & $0.8175_{-0.0093}^{+0.0093}$                  \\ 
$\nu$             & 0                             & $(8.372\times10^{-05})_{-0.00018}^{+0.00018}$ \\ \hline\hline
\multicolumn{3}{c}{CMB+BAO+SN IA+RSD analysis}                                                    \\ \hline\hline
Parameter         & $\Lambda$CDM                  & RGGR                                          \\ \hline
$100~\omega_{b }$ & $2.242_{-0.014}^{+0.013}$     & $2.241_{-0.015}^{+0.015}$                     \\ 
$\omega_{cdm }$   & $0.1191_{-0.00095}^{+0.0009}$ & $0.119_{-0.00095}^{+0.001}$                   \\ 
$n_{s }$          & $0.9662_{-0.0036}^{+0.0038}$  & $0.967_{-0.0044}^{+0.0042}$                   \\ 
$\tau_{reio }$    & $0.05364_{-0.0074}^{+0.007}$  & $0.0535_{-0.0069}^{+0.0069}$                  \\ 
$H_0$             & $67.77_{-0.41}^{+0.42}$       & $67.86_{-0.47}^{+0.46}$                       \\ 
$\sigma_8$        & $0.8068_{-0.0057}^{+0.0059}$  & $0.8089_{-0.0087}^{+0.0081}$                  \\ 
$\nu$             & 0                             & $(5.815\times10^{-05})_{-0.00017}^{+0.00017}$ \\ \hline\hline
\end{tabular}
\label{tab:results}
\end{table}

\section{Conclusions} \label{conclusions}

In this work, we have explored the cosmological implications of a gravity model, fully defined from an action, where both the gravitational constant ($G$) and the cosmological constant ($\Lambda$) exhibit scale dependence at perturbative levels, while preserving the background cosmological field equations. This scale dependence arises from the application of renormalization group effects to the Einstein-Hilbert action, as outlined in \cite{Shapiro:2009dh, Shapiro:2004ch, Reuter:2003ca, Reuter:2004nx, Rodrigues:2009vf, Domazet:2010bk, Rodrigues:2015hba} (and references therein). The precise cosmological framework here considered, with two scales and a complete action, extends the previous works and was introduced by some of us in Ref.~\cite{Bertini:2019xws}. We refer to this general framework as Renormalization Group improved General Relativity (RGGR) (this name was introduced in Ref.~\cite{Rodrigues:2009vf}). Here we are concerned with the scale dependence induced by such approaches and their large scale classical consequences.

Within this two-scale RGGR approach, a single parameter $\nu$ models GR deviations (with $\nu=0$ corresponding to GR). In Ref.~\cite{Bertini:2019xws}, slip and $f \sigma 8$ observational constraints were considered, and it was found that $|\nu| \lesssim 10^{-1}$, with a clear tendency toward negative $\nu$. In the present work, one sees that the qualitative analysis of the CMB acoustic oscillations (Fig.~\ref{fig:cmb}) is sufficient to find stronger constraints $\nu$. The use of more recent and diverse observational data, including CMB data from Planck 2018 and the Pantheon Type Ia supernovae dataset, allowed us to constrain the value of $\nu$ to $|\nu| \lesssim 10^{-5}$. This implies a small deviation from the standard $\Lambda$CDM model, in particular as shown in Fig.~\ref{plotplanckbaorsd} and Table \ref{tab:results}. From the latter, the cosmological parameter most susceptible to a change in $\nu$ from zero is found to be $\sigma_8$, but the change in its mode or its maximum 2$\sigma$ value is about or smaller than $0.5\%$. Hence, this model is not suitable to answer the current cosmological issues. The  $\nu$-parameter does not provide a statistically significant improvement in fitting the data compared to the $\Lambda$CDM model. The very small value of $\nu$ suggests that its impact on cosmology is negligible. Consequently, one can consider that the RGGR, as implemented here, is disfavored, as the $\Lambda$CDM model explains the observational data more simply with fewer parameters.

In this work, we used a modified CLASS Boltzmann solver together with MontePython for the statistical analysis. Despite the small $\nu$ value obtained, we presented here a viable framework for incorporating scale-dependent effects in cosmology,  offering new insights into the discrepancies observed in acoustic oscillations and other cosmological measurements. This approach is theoretically consistent and may have other phenomenological advantages, but at the moment its differences with respect to GR are not relevant to address the $H_0$ or $\sigma_8$ issues that $\Lambda$CDM faces. Different approches with renormalization-group  motivation and impact on these issues can be found, for instance, in Refs.~\cite{DiValentino:2021izs, SolaPeracaula:2021gxi}. Further investigations into the RGGR framework, perhaps in line with the parallel development of \cite{Bertini:2024onw}, could provide new explanations for puzzling cosmological phenomena. 
\bigskip

\noindent
\textbf{Acknowledgements}
The authors thank Ilya Shapiro for relevant discussions on gravity and renormalization group effects. WSHR acknowledges \textit{Fundação de Amparo à Pesquisa e Inovação do Espírito Santo}  (FAPES, Brazil) for partial support. RvM is suported by \textit{Funda\c{c}\~ao de Amparo \`a Pesquisa do Estado da Bahia} (FAPESB) grant TO APP0039/2023. FMS acknowledges FAPES for support. DCR acknowledges \textit{Conselho Nacional de Desenvolvimento Científico e Tecnológico} (CNPq, Brazil) and FAPES (Brazil) for partial support. The authors acknowledge the use of the computational resources provided by the Sci-Com Lab of the Department of Physics at UFES, which is funded by FAPES and CNPq. 


\bibliographystyle{unsrtnat}
\bibliography{references}

\appendix

\section{Initial conditions} \label{initialconditions}
To set the initial conditions of the Boltzmann- Einstein equations system, the limit at a given initial time $ \eta_i > 0 $, in very large scales  ( $ k \eta \ll 1 $) and  deep in the  radiation dominated era ( $ a \propto \eta $)  are solved. To this it is commonly done a series expansion  with respect to $k\eta$ in all perturbative quantities, i.e. $ Y = \Sigma^{\infty}_{n = 0} Y^{(n)} (k \eta)^{n} $, where $Y$ is any relevant perturbative quantity.

Since photons, neutrinos and baryons equations do not change for the RGGR case,   the relevant limits for those species  remain the same as GR. For example  in the case of photons,  multipolar moments of Bolztmann equations for $l \geq 2$ are negligible, while the same happens for the neutrinos case
for $l > 2$, which implies that only monopoles ($\delta_\gamma$, $\delta_n$), dipoles ($V_\gamma$, $V_n$)  and the neutrino quadrupolar moment $\mathcal{N}_2$ are relevant. The  limits  for photons, neutrinos and baryons contrast density equations lead us to \cite{Piattella:2018hvi,Ma:1995ey}
\begin{eqnarray} \label{iniOthers}
    \delta'_\gamma = 4 \phi' \,,  \qquad  \delta'_n = 4 \phi' \,,   \qquad  \delta'_b = 3 \phi'\,,
\end{eqnarray}
and for velocities \cite{Piattella:2018hvi,Ma:1995ey}
\begin{eqnarray}\label{velocity1}
    V_\gamma = V_n + q_n = V_b= V_c = \frac{k\eta}{2}\psi\;, 
\end{eqnarray}
where $V_b$ is the baryons velocity and $q_n$ is at most a linear function of $\eta$ and is the neutrino velocity isocurvature mode or relative neutrino heat flux. The last equality above is because the $Q^\beta$-term only affects the energy transfer and not the moment transfer (see Eq.~(\ref{cdmV})).

In the case of dark matter,  the $\delta_c $, $ V_c $, $ \phi $ and $ \psi $  expansion series are introduced in Eq.(\ref{cdm}) and in relevant limits we have
\begin{eqnarray} \label{iniDM}
    \delta'_{c} = 3 \phi' + \frac{6 \nu}{\eta} \psi \,,
\end{eqnarray}
Integrating eqs.(\ref{iniOthers}) and (\ref{iniDM}) we get 
\begin{eqnarray} \label{solutions}
    \delta_\gamma &=& 4 \phi + 4 C_\gamma \,, \qquad 
    \delta_n =  \delta_\gamma + S_n \,, \qquad  \nonumber \\ 
    \delta_b &=&\frac{3}{4}\delta_\gamma + S_b \, 
      \qquad  \;\;\delta_c = \frac{3}{4}\delta_\gamma  + S_c + 2 f(\eta)\;,
\end{eqnarray}
with constants in $\eta$ 
\begin{eqnarray}
    S_n \equiv C_n - 4 C_\gamma\;,\;\;\; S_c \equiv C_c - 3 C_\gamma\;\;\; \mbox{e} \;\;\;S_b \equiv C_b - 3 C_\gamma\;,
\end{eqnarray}
and
\begin{eqnarray} 
    f(\eta) = 3 \nu \int  \psi (\eta)  d \ln  \eta \,. \label{eff}
\end{eqnarray}
$S_n$, $S_c$ and $S_b$ are the neutrino density isocurvature modes, CDM density isocurvature modes and baryons density isocurvature modes respectively. However here we shall consider adiabatic perturbations in which $S_n=S_c=S_b=q_n=0$ and  $C_\gamma \neq 0$ and hence, from eq.(\ref{solutions}) 
\begin{eqnarray}
    \frac{1}{3} \delta_c -  \frac{2}{3}f(\eta) = \frac{1}{3} \delta_b = \frac{1}{4} \delta_\gamma = \frac{1}{4} \delta_n = \phi + C_\gamma \label{adiatmodesdelta} \,.
\end{eqnarray}
It is well known that in the adiabatic case, $C_\gamma$  is equal to the curvature perturbation coming from inflation,  and   also it is constant on large scales. To find the initial conditions for   density contrasts and velocities, it is necessary to known expressions for potentials $\psi$ and $\phi$.  Eq. (\ref{poisson})   in  the limit  $ k \eta \ll 1 _i$ takes the form
\begin{eqnarray} 
     - 2 \eta \phi'  - 2 \psi - 2 \nu \eta \psi' - 2 \nu \psi  = R_\gamma \delta_\gamma +R_b \delta_b+R_n \delta_n+R_c \delta_c \;, \label{poissonprev}
\end{eqnarray}
where the density fraction was defined as $ R_i \equiv \frac{\tilde{\epsilon}_{i}}{\tilde{\epsilon}}$ for each species, with condition
$ R_\gamma + R_n + R_c + R_b = 1 $. In the radiation dominated era $R_\gamma + R_n \approx 1$ and $R_c$ and $R_b$ are negligible. With this in mind,  solutions (\ref{solutions}) are introduced in (\ref{poissonprev}) and for the adiabatic case we have
\begin{eqnarray} \label{poissonCIAd}
     -  \eta \phi' -2 \phi -  (1+\nu) \psi -  \nu \eta \psi'  = 2 C_\gamma \;.  
\end{eqnarray}

On the other hand, in the relevant limits, the term on the right of Eq. (\ref{traceless}) has  only the  neutrinos quadrupolar term contribution \cite{Ma:1995ey}, i.e  ($\sigma = 2 \mathcal{N}_2$). Thus  Eq. (\ref{traceless}) takes the form 
\begin{eqnarray} \label{quadrupoloneutrinoCIAd}
    (k\eta)^2(-\phi + (1-2\nu) \psi) = -12 R_n \mathcal{N}_2\;. 
\end{eqnarray}
Finally,  pressure perturbations equation (\ref{pressure}) in the relevant limits and for the adiabatic case is written as
\begin{eqnarray} \label{pressureLimit}
     4 \eta \phi' + 2 \eta^2 \phi'' - 2 (1  + \nu) \psi + 2 (1  + \nu) \eta \psi' + \;\;\;\;\;\;\;\;\;\;\; \;\;\;\;\nonumber \\ 2 \nu \eta^2 \psi'' =   
      R_\gamma \delta_\gamma +R_n \delta_n \,.
\end{eqnarray}
in which solutions (\ref{solutions})  can be introduced. For the adiabatic case this lead to
\begin{eqnarray}\label{pressureCIAd}
    -2 \phi  + 2 \eta \phi' +  \eta^2 \phi'' -  (1  + \nu) \psi +  (1  + \nu) \eta \psi' + \;\;\;\;\;\;\;\;\;\;\; \;\;\;\;\nonumber \\   \nu \eta^2 \psi'' = 2 C_\gamma\;.
\end{eqnarray}

From eqs.(\ref{poissonCIAd}), (\ref{quadrupoloneutrinoCIAd}) and (\ref{pressureCIAd}) it is possible to find a fifth-order differential equation for the $\phi$ potential
\begin{eqnarray} \label{phiCI}
    -\nu \eta^4 \phi'''''+ (1-18 \nu ) \eta^3 \phi'''' + 12 \left[ 1 - \nu \left(8 + \frac{2R_n}{15}\right)\right] \eta^2 \phi''' \nonumber  \\ +4  \left[9 + \frac{2}{5} R_n- \left(42 + \frac{12}{5} R_n\right) \nu\right] \eta \phi''  \nonumber\\
      + 8 (1-3 \nu ) \left(3 + \frac{2}{5} R_n\right) \phi' = 0\, 
\end{eqnarray}
where the terms of the order $ \nu^2 $ and higher were neglected. 

We must emphasize that the $\nu$-dependence in $\delta_c$, $\phi$ and $\psi$ equations leads to differences with GR adiabatic initial conditions. This dependence is inherited by all initial conditions for the other species. The case $\nu =0$  reduces to the GR initial conditions. Solutions for eq. (\ref{phiCI}) have the form
\begin{eqnarray}
   \phi &\propto& \eta^{ -\frac{5}{2} \pm \frac{\sqrt{5-32 R_n}}{2 \sqrt{5}}} \;\;,\qquad \phi \propto \frac{1}{\eta} \;\;,\nonumber \\ 
  \phi &\propto& \frac{\eta^{\frac{1}{\nu } - 2}}{\frac{1}{\nu}- 2} \;\;\;\;\;\;\;\;\;\;\;\;,\qquad \phi = \mbox{constant}= \phi_0
\end{eqnarray}
The first two solutions are complex  when $ R_n > \frac{5}{32} $ thus we shall focus only in the cases with $ R_n <\frac{5}{32} $, but in those cases the solutions are decaying, then we shall not consider them. The same happens for the third solution i.e it is also a decaying solution. The fourth solution is also not considered   because it has one  indetermination in the limit when $ \nu = 0 $. However our objective here is to study solutions that have a  $ \Lambda $CDM limit. Then the only useful solution is the last one i.e $\phi = \mbox{constant}= \phi_0$, which implies that $\psi = \mbox{constant}= \psi_0$.  Derivating twice Eq.(\ref{quadrupoloneutrinoCIAd}), 
we have
\begin{eqnarray}
    2  \left[ - \phi + (1-2\nu)\psi\right] + 4  \eta \left[-\phi' + (1-2\nu)\psi'\right]  +  \nonumber \\ \eta^2\left[-\phi'' + (1-2\nu)\psi''\right]
     = - \frac{8}{5} R_n  \frac{V'_n}{k}= - \frac{8}{5} R_n \left( \psi + \frac{\delta_n}{4} \right)\;,\label{primeiraderiveqij}&&
\end{eqnarray}
where we used GR relations for neutrinos, $5\mathcal{N}'_2 = 2 \,k \, \,\mathcal{N}_{1} =2 \,k \, V_n/3$. 
Derivating again and using the fact that $\delta'_n = 4 \phi'$ we have
\begin{eqnarray}
    6 \left[-\phi' + (1-2\nu)\psi'\right] + 6 \eta \left[-\phi'' + (1-2\nu)\psi''\right]  +  \nonumber \\ \eta^2\left[-\phi''' + (1-2\nu)\psi'''\right] = - \frac{8}{5} R_n \left( \psi' + \phi' \right)\;.&&  \label{segundaderiveqij}
\end{eqnarray} 
On the other hand, from equation (\ref{poissonCIAd}) is possible to write
\begin{eqnarray}
    \eta \psi' = \left[- 2 \phi - 2  \eta \phi' - (1 + \nu)\psi  - 2 C_\gamma \right] \nu^{-1} \;, \nonumber 
\end{eqnarray}
which replacing in (\ref{pressureCIAd}) and neglecting the order $ \nu^2 $ and highers, helps to find a relation between $\psi$  and $\phi$ and its derivatives
\begin{eqnarray} 
    \psi = \nu \eta^2 \phi'' - (1 - 4\nu) \eta \phi' - 2 (1-\nu) \phi  - 2(1-\nu) C_\gamma\;.  \label{PsiCI}
\end{eqnarray}

From the equations (\ref{primeiraderiveqij})  and (\ref{PsiCI})  for $\phi_0$
 and  $\psi_0$  up to first order of $\nu$  we find the system
 \begin{eqnarray} \label{systemPhiPSi}
        &&-2 (1-\nu) \phi_0 - \psi_0 =  2(1-\nu) C_\gamma \nonumber\\
    &&- \phi_0 + (1-2\nu)\psi_0  = - \frac{4}{5} R_n \left( \psi_0+ \phi_0 + C_\gamma \right) \,.
\end{eqnarray}
Solutions for the system (\ref{systemPhiPSi}) are
\begin{eqnarray}
    \phi_0 &=& -\frac{2 C_\gamma (5 +2 R_n )}{15 + 4 R_n} \left[ 1 - \frac{5 \nu}{5 +2 R_n}\right] \;\;\;, \qquad \nonumber \\
    \psi_0 &=& -\frac{10 C_\gamma}{15 + 4 R_n}(1+\nu)\,.
\end{eqnarray}
From (\ref{adiatmodesdelta}) we can find the initial conditions for the density contrast for all species
\begin{eqnarray}
    \delta_\gamma = \delta_n= \frac{4}{3}\delta_b= \frac {20 C_\gamma} {(15+4 R_n)}  (1+2\nu) \,.
\end{eqnarray}

Now we can integrate (\ref{eff}) to find $f(\eta)$ function and $\delta_c$ up to first order of $\nu$
\begin{eqnarray}
    f(\eta) & =& 6 \nu \, \psi_0 \,\ln \left(\frac{\eta}{\eta_0} \right)\;\;\;, \qquad  \nonumber \\ \delta_c &=&\frac{15 C_\gamma}{15+4R_n}\left[1+2\nu\left(1-2\ln \left(\frac{\eta}{\eta_0}\right)\right)\right] \,.
\end{eqnarray}
where $\eta_0$ is the integration constant. In our numerical analysis we consider $\eta \sim \eta_0$, therefore, logarithmic term is canceled. Initial conditions for velocities are given by eqs.(\ref{velocity1}) but with $\psi=\psi_0$. Finally from (\ref{quadrupoloneutrinoCIAd}), (\ref{primeiraderiveqij}) and (\ref{PsiCI}), neglecting terms of second order and higher in $\nu$, we  calculate the primordial mode of the neutrino quadrupolar moment as
\begin{eqnarray}
    \mathcal{N}_2 = \frac{k^2\eta^2}{30}\,\psi_0 (1-\nu)\;.
\end{eqnarray}

\end{document}